\documentclass{moriond}

\def\be{\begin{equation}}
\def\ee{\end{equation}}
\def\bea{\begin{eqnarray}}
\def\eea{\end{eqnarray}}

\usepackage{amsmath,amssymb}
\usepackage{bm}
\usepackage{graphicx} 
\usepackage{hyperref}
\usepackage{mathrsfs}
\usepackage{multirow}    
\usepackage{dsfont}    
\usepackage{bbm}

\usepackage{ulem}
\usepackage[usenames]{color}

\allowdisplaybreaks

\newcommand{\Photo}{}
\newcommand{\ud}{\mathrm{d}}

\begin{document}
\vspace*{4cm} \title{HIGH-ORDER COMPARISONS BETWEEN \\POST-NEWTONIAN
  AND PERTURBATIVE SELF FORCES}

\author{Luc BLANCHET and Guillaume FAYE}

\address{$\mathcal{G}\mathbb{R}\varepsilon{\mathbb{C}}\mathcal{O}$
  Institut d'Astrophysique de Paris, UMR 7095, CNRS,\\ Sorbonne
  Universit{\'e}s \& UPMC Univ Paris 6, 98\textsuperscript{bis}
  boulevard Arago, 75014 Paris, France}

\author{Bernard~F. WHITING}

\address{Institute for Fundamental Theory, Department of Physics,\\
  University of Florida, Gainesville, FL 32611, USA}

\maketitle\abstracts{Recent numerical and analytic computations based on the
  self-force (SF) formalism in general relativity showed that half-integral
  post-Newtonian (PN) terms, i.e. terms involving odd powers of $1/c$, arise
  in the redshift factor of small mass-ratio black-hole binaries on exact
  circular orbits. Although those contributions might seem puzzling at first
  sight for conservative systems that are invariant under time-reversal, they
  are in fact associated with the so-called non-linear tail-of-tail effect. We
  shall describe here how the next-to-next-to-leading order contributions
  beyond the first half-integral 5.5PN conservative effect (i.e. up to order
  7.5PN included) have been obtained by means of the standard PN formalism
  applied to binary systems of point-like objects. The resulting redshift
  factor in the small mass-ratio limit fully agrees with that of the SF
  approach.}

\section{Introduction}

Stellar-mass compact objects inspiraling gradually about super-massive black
holes may produce gravitational waves detectable by future space missions such
as eLISA~\cite{Whitepaper}. These systems, referred to as Extreme Mass Ratio
Inspirals (EMRIs), can probe the strong gravity field regime, but proper data
analysis of the resulting signal will require accurate waveform templates
built from theoretical models. This has motivated, over the past ten years,
numerous studies on the dynamics of point-like objects on a curved
background~\cite{MiSaTa,QuWa,PoissonLR}. Due to the metric perturbations
generated by its own mass-energy, the point mass effectively feels a
self-force (SF) that induces deviations from the geodesic wordline followed by
a test particle on the background. When the background is a black-hole
spacetime, the acceleration may be sought in the form of an asymptotic
expansion in powers of the mass ratio $q=m_1/m_2 \ll 1$.

In the SF approach, the first order perturbation $\delta g_{\mu\nu}$ of the
background metric $g_{\mu\nu}^{(0)}$ is obtained by convolving over the
stress-energy tensor $T^{\mu\nu}$ the regularized (R) Green function
$G_{\text{R}~\alpha'\beta'}^{\mu\nu}(x, x')$ that solves the linearized
homogeneous Einstein equations in harmonic gauge and has the property that it
coincides with the retarded Green function when $x$ lies in the chronological
future of $x'$, while vanishing when $x$ is in the chronological past of
$x'$~\cite{DW03}. The trajectory of the particle is then precisely that of a
geodesic for the perturbed metric $g_{\mu\nu}^{(0)} + \delta g_{\mu\nu}$.

By contrast, the post-Newtonian (PN) approach is based on the formal
expansion, on a flat background spacetime, of all quantities of interest, in
powers of $v/c$, where $v$ is the largest typical velocity of the problem. The
standard PN approach is first defined for general extended PN sources with
compact support and then specialized to compact binary
systems~\cite{Bliving14}. In that case, $v$ is taken to be the relative
coordinate velocity ${v}_{12}$. Moreover, when the bodies are compact, they
may be effectively represented as point particles. Ultra-violet (UV)
divergences at the particle positions are tackled by means of dimensional
regularization~\cite{tHooft,Bollini}. The PN expressions are valid at a given
coordinate time in a spatial region, referred to as the near zone, that
entirely contains the matter source and whose radius is much smaller than the
gravitational wavelength. Because SF and PN methods are so radically different
from each other, notably regarding the regularization schemes, comparing PN
expansions of observable quantities truncated at linear order in $q$ to their
SF counterparts expanded in power of $1/c$ allows for non-trivial cross-checks
that strengthen our confidence in both perturbative techniques.

After the first comparison between PN and gravitational SF
calculations~\cite{Det08}, rapid progress has been made over the last six
years, mainly due to both high precision numerical computations from a SF
perspective and extensive analytical PN computations~\cite{BDLW10a,BDLW10b}.
Recently, after the possibility for this comparison had been dramatically
extended from the SF side~\cite{SFW14}, it was realized that observable
quantities could contain half-integral $\frac{n}{2}$ PN terms that are
nevertheless conservative, starting at the order 5.5PN~\cite{SFW14}. Here, we
shall explain how those terms can arise within the PN framework~\cite{BFW14a}
and sketch their calculation at the next-to-next-to-leading
order~\cite{BFW14a,BFW14b}. As we shall see, they are closely related to the
so-called tail-of-tail effects in general relativity. The success of this
SF/PN comparison actually provides an excellent test of the intricate PN
machinery for computing non-linear tail-of-tail effects --- these being
relevant for template waveform generation of comparable mass compact binaries
to be analyzed in ground and space based detectors.

\section{Comparing post-Newtonian and self-force results}

\subsection{The Detweiler variable}

We shall focus on the Detweiler variable~\cite{Det08}, which represents
physically the inverse of the redshift of a photon emitted by a particle
moving on an exact circular orbit around a Schwarzschild black hole and
detected by an infinitely far-away observer along the rotation axis. The
ensuing spacetime is helically symmetric, with a helical Killing vector
$K^\alpha$ tangent to the four-velocity $u_1^\alpha$ on the particle
worldline. Alternatively, the redshift variable is defined geometrically as
the conserved quantity associated with the helical Killing symmetry relevant
to spacetimes with exactly circular orbits. In an appropriate class of
coordinate systems, the redshift factor $u_1^T$ reduces to the $t$ component
of the particle's four-velocity,
\begin{equation}\label{uTdef}
u_1^T = \frac{1}{\sqrt{- g_{\alpha\beta}(y_1) v_1^\alpha v_1^\beta/c^2}}\,,
\end{equation}
where $g_{\alpha\beta}(y_1)$ denotes the metric evaluated at the particle's
location $y_1^\alpha = (c t, y_1^i)$ by means of dimensional regularization,
and $v_1^\alpha\equiv\ud y_1^\alpha/\ud t=(c,v_1^i)$ is the coordinate
velocity.

The Detweiler variable~\eqref{uTdef} has been computed to high-order using on
the one hand standard PN theory suplemented with dimensional regularization,
valid in weak field~\cite{BDLW10a,BDLW10b}, and on the other hand both
numerical and analytical SF approaches, valid in the limit $q=m_1/m_2 \ll 1$.
Over the last two years, its accuracy has improved drastically on the SF side
due to the new application of methods to represent analytic solutions for
metric perturbations of black-hole spacetimes. In a first stage, based on some
exact solutions of the Teukolsky equation~\cite{MST96a}, the PN coefficients
of the redshift factor were obtained numerically to 10.5PN order~\cite{SFW14}.
Analytic expressions were even found for a subset of coefficients,
specifically those that are either rational, or made of the product of $\pi$
with a rational, or a sum of commonly occurring transcendentals. An
alternative SF approach~\cite{BiniD14b}, based on the post-Minkowskian
expansion of the Regge-Wheeler-Zerilli equation~\cite{MST96b}, has also
reached PN coefficients analytically up to 8.5PN order. Most recently, both
methods have been extended to extremely high orders, typically 21.5PN for the
redshift factor~\cite{KOW15,McDSW15}.

The appearance of \textit{half-integral} PN coefficients (of type
$\frac{n}{2}$PN with $n$ being an odd integer) in the conservative dynamics of
two particles on circular orbits is a feature of high-order PN expansion.
Resorting to standard PN methods, we shall show now that the leading
half-integral PN terms originate from non-linear integrals depending on the
past history of the source --- so-called hereditary type
integrals.

\subsection{Dimensionality argument}
\label{sec:dimarg}

The fact that terms at half-integral PN orders cannot stem from the
source variables evaluated at the current time follows from the
general structure of ``instantaneous'' terms entering the redshift
factor~\eqref{uTdef} in the center-of-mass frame. After replacing all
accelerations by the lower-order equations of motion, $u_1^T$ takes
the form (with usual Euclidean notation)
\begin{equation}\label{uTinst}
\left(u_1^T\right)_\text{inst} \sim \sum_{j, k, n, p\atop\text{integers}}
\,\nu^j\left(\frac{G m}{r_{12}
  c^2}\right)^k\left(\frac{\bm{v}_{12}^2}{c^2}\right)^n
\left(\frac{\bm{n}_{12}.\bm{v}_{12}}{c}\right)^p\,,
\end{equation}
where $m=m_1+m_2$, $\nu=m_1m_2/m^2$, while
$\bm{n}_{12}=(\bm{y}_1-\bm{y}_2)/r_{12}$ stands for the relative direction of
the two particles. We have taken the expansion when the mass ratio $\nu\to 0$.

The counting of the $1/c$ powers shows that the PN order of the generic term
in Eq.~\eqref{uTinst} can be half-integral only if $p$ is odd, in which case
it vanishes for circular orbits, when the velocity $\bm{v}_{12}$ and unit
direction $\bm{n}_{12}$ are evaluated \textit{at the same current time} $t$.
However, integration over some intermediate time extending from the infinite
past up to $t$ could allow a coupling between these vectors at different
times. Of course, in general relativity, this type of ``hereditary''
dependence over the past of the system does occur, due, in particular, to wave
tails produced by the back-scatter of linear waves on the spacetime curvature.

\section{Post-Newtonian computation of half-integral order contributions}

\subsection{Structure of the tail contributions}

The tail effects, associated with non-linear wave propagation, are best
investigated by constructing the (multipolar-)post-Minkowskian expansion of
the metric $g_{\mu\nu}$ in powers of the gravitational constant $G$, outside
the matter source~\cite{Bliving14}. We start from the Einstein field equations
in vacuum, written in terms of the field perturbation variable
$h^{\mu\nu}\equiv\sqrt{-g}g^{\mu\nu}-\eta^{\mu\nu}$ on the flat background
$\eta^{\mu\nu}$ in Cartesian coordinates $\{x^i\}$, with $g$ representing the
determinant of $g_{\mu\nu}$. Adopting the harmonic-gauge condition
$\partial_\nu h^{\mu\nu} =0$, the relaxed Einstein equations for $h^{\mu\nu}$
reduce to the wave-like equations $\Box h^{\mu\nu} = \Lambda^{\mu\nu}$, with
$\Box\equiv\eta^{\alpha\beta}\partial_\alpha\partial_\beta$. The non-linear
source term $\Lambda^{\mu\nu}$ is an expression of second-order (at least) in
$h^{\alpha\beta}$. At linear order in $G$, the most general solution depends
on six sets of source multipole moments: the mass-type moments, $I_L\equiv
I_{i_1...i_\ell}$ ($\ell$ being the multipole order), the current-type moments
$J_L\equiv J_{i_1...i_\ell}$, and four sets of so-called gauge moments,
irrelevant for the present discussion. The higher order solutions are obtained
iteratively by applying the flat retarded integral operator
$\Box^{-1}_\text{ret}$ on the source term, after multiplication by a
regularization factor $r^B$ to cope with the divergence of the multipole
expansion when $r\equiv|\bm{x}|\to 0$. Analytic continuation in
$B\in\mathbb{C}$ is invoked and the finite part (FP) when $B\to 0$ provides a
certain particular solution. To ensure that the harmonic coordinate condition
is satisfied at each step, we must add to the latter solution a specific
homogeneous retarded solution, which does not generate tail integrals and can
be safely ignored here.

Since, ultimately, we shall be interested in the metric at the location of one
of the particles, our goal is to compute the near-zone expansion, indicated
below by an overline, of the general solution initially defined in the
exterior of the source, when $r\to 0$. It is obtained directly at a given
order from the near-zone expansion of the corresponding source, without need
to control the full solution, from the formula~\cite{B93}:
\begin{align}
 \overline{\mathop{\mathrm{FP}}_{B=0}\Box^{-1}_\text{ret}
   \bigl[\hat{n}_L S(r,t-r/c)\bigr]} =
 \hat{\partial}_L\left\{\frac{\overline{\mathcal{G}(t-r/c)
     -\mathcal{G}(t+r/c)}}{r}\right\} +
 \mathop{\mathrm{FP}}_{B=0}\Box^{-1}_\text{inst}\bigl[\hat{n}_L
   \overline{S(r,t-r/c)}\bigr] \,,\label{sol0}
\end{align}
where $\hat{n}_L$ denotes the symmetric trace-free part of $n_L\equiv
x^{i_1}\!...\, x^{i_\ell}/r^\ell$ ($\ell \in \mathbb{N}$). The first
term is a homogeneous solution of the wave equation which is of
retarded-minus-advanced type and regular at $r= 0$. It may be directly
expanded in the near-zone, where it is valid by virtue of a matching
argument. The second term in Eq.~\eqref{sol0} is a particular solution
of the inhomogeneous equation that is defined by means of the
``instantaneous'' inverse box operator, $\Box^{-1}_\text{inst}$,
representing the PN expansion of the symmetric integral operator
supplemented with a regulator $r^B$ multiplying the source and a
finite part as $B\to 0$. This term diverges when $r\to 0$ and should
be matched to a full-fledge solution of the field equations inside the
source. However, we proved~\cite{BFW14a} (see the appendix there) that
it cannot actually contribute at half-integral PN orders, so that the
effect we are looking for comes only from the term containing
$\mathcal{G}(u)$. The latter function is given by a specific double
integral over the source piece $S(r,t)$, and regularized by the finite
part as $B\to 0$,
\begin{align}
\mathcal{G}\left(u\right) &=
\mathop{\mathrm{FP}}_{B=0}\int_{-\infty}^u \ud
s\,\mathop{R}_B\left(\frac{u-s}{2},s\right)\,,\label{G}\\ \text{where}\quad
\mathop{R}_B\left(\rho,s\right) &= 2^{\ell-1}\rho^\ell \int_0^\rho\ud
\lambda\,\frac{(r-\lambda)^\ell}{\ell!}\lambda^{B-\ell+1}
S(\lambda,s)\,.\label{R}
\end{align}
This function is the crucial object to investigate for the purpose here.

At linear order in the mass ratio, we may disregard any hereditary term
involving the product of more than two moments other than the mass monopole
$M$, since each such multipole is proportional to $\nu$. General results on
the structure of the gravitational field in harmonic gauge tell us that
hereditary contributions of type $M\times\cdots\times M \times \mathcal{M}_P$,
with $\mathcal{M}_P=I_P$ or $J_P$, read~\cite{Bliving14}
\begin{equation}\label{hstruct}
h^{\mu\nu}_{M \times\cdots\times M \times \mathcal{M}_P} \sim
\sum_{k,p,\ell,i\atop\text{integers}} \frac{G^{k}
  M^{k-1}}{c^{3k+p}}\,\hat{n}_L\left(\frac{r}{c}\right)^{\ell+2i}
\int_{-\infty}^{+\infty}\ud u\,\kappa^{\mu\nu}_{LP}(t,u)\,
\mathcal{M}^{(a)}_P(u)\,,
\end{equation}
where the upper sign $(a)$ refers to time derivatives and the tensor function
$\kappa^{\mu\nu}_{LP}(t,u)$ is a dimensionless kernel. Using dimensional
analysis combined with ``angular-momentum'' selection rules, it is
straightforward to show that only interactions with $k\geqslant 3$ can produce
the half-integral PN terms of interest, starting at the leading 5.5PN order.
In fact, at the next-to-next-to-leading order beyond 5.5PN, we can restrict
ourselves to hereditary cubic interactions $M\times M \times \mathcal{M}_P$,
which may be interpreted physically as gravitational-wave tails of tails. They
must be computed at the 5.5PN, 6.5PN and 7.5PN orders for the mass quadrupole,
6.5PN and 7.5PN orders for the mass octupole and current quadrupole, and so
on. The leading contributions of current-type moments are 1PN order higher
than those of the mass moments.

\subsection{Sketch of the calculation of $u^T_1$}

The elementary source terms $\hat{n}_L S(r,t-r/c)$ for the tails of tails can
be either instantaneous, with $S(r,t-r/c)=r^{k} \mathcal{M}^{(a)}_P(t-r/c)$,
or hereditary. In the latter case, $S(r,t-r/c)$ is an integral of the type
$r^{-k} \int_1^{+\infty} \!\ud x \, Q_m(x) \mathcal{M}^{(a)}_P(t-r x/c)$, with
$Q_m(x)$ being a Legendre function of the second kind. After some
transformations, the tail-of-tail piece of $\mathcal{G}(u)$ can be written as
the finite part at $B=0$ of some coefficient $C_{\ell k m}(B)$, times an
integral of the variable $\tau$ whose integrand involves the regulator
$\tau^B$ times derivatives of $\mathcal{M}_P(t-\tau)$. Now, we find that
$C_{\ell k m}(B)$ may comprise (simple) poles at the order we are working, so
that the factor $\tau^B$ generates a logarithm kernel $\ln \tau$. Insertion of
the former piece of $\mathcal{G}$ into the homogeneous wave entering
Eq.~\eqref{sol0} yields the form~\eqref{hstruct}, with
$\kappa^{\mu\nu}_{LP}(t,u)$ proportional to $\ln (t-u)$ for $t>u$, and zero
elsewhere~\cite{BFW14a,BFW14b}.

At this stage, it is important to realize that this sort of ``pure''
tail-of-tail contributions can generate another kind of half-integral terms,
``mixed'' contributions, by coupling to the non-tail part of the PN metric in
the source of Einstein's equations. This part is obtained most conveniently by
solving the relaxed field equations in the near zone, where $\Lambda^{\mu\nu}$
is augmented by the matter source $16\pi Gc^{-4}|g| T^{\mu\nu}$. It is usually
parametrized by means of appropriate potentials, such as the Newtonian
potential $U = G m_1/r_1+G m_2/r_2$ (with $r_1=|\bm{x}-\bm{y}_1|$). The time
component $h^{00}$ of the gravitational field, for instance, is composed of
``ordinary'' PN terms: $-4U/c^2+\cdots$, plus tail terms containing our
effect: $h^{00}_\text{\text{tail}~(5.5\text{PN})}+\cdots$. Its product with,
say, $h^{ij}$, whose structure is similar, produces couplings that must be
crucially taken into account in the calculation. Their number is minimized by
moving to an adapted gauge. Quadratic and cubic PN iterations are then
required to find the complete half-integral PN part of the metric at the 7.5PN
order. The successive solutions are constructed by means of hierarchies of
``superpotentials'' derived from the potentials that enter $g_{\mu\nu}$ at the
2PN order~\cite{BFW14b}. In the end, we decompose the tail integrals into
conservative time symmetric and dissipative time anti-symmetric pieces and
simply discard the dissipative piece from our results.

Using the standard stress-energy tensor for point particles to model the
binary, all multipole moments, potentials and superpotentials can be evaluated
explicitly for circular orbits. The hereditary integrals are derived from
standard formulas. This yields for the redshift~\cite{BFW14a,BFW14b}
\begin{equation}\label{uTSF}
u_\text{SF}^T = - y - 2 y^2 -5 y^3 +... -
\frac{13696}{525}\pi\,y^{13/2} + \frac{81077}{3675}\pi\,y^{15/2} +
\frac{82561159}{467775}\pi\,y^{17/2} + ...\,,
\end{equation}
where we have written only the relative 2PN terms relevant to our
next-to-next-to-leading order calculation, with all the other terms indicated
by ellipsis. The result~\eqref{uTSF} is in full agreement with those derived
by gravitational SF methods, either semi-analytical or purely
analytical~\cite{SFW14,BiniD14b}. We emphasize that it has been achieved from
the standard PN approach, which is not tuned to a particular type of source
(contrary to various analytical and numerical SF calculations), as it is
actually applicable to any extended PN source with compact support. The 7.5PN
order reached by the present calculation is perhaps the highest order ever
reached by traditional PN methods. Note also that while SF results may be
relativistic but with $q\ll 1$, the present method, valid primarily in the PN
regime, is in principle applicable for arbitrary mass ratios. The time is now
ripe for the SF approach to proceed to second order in $q$. We look forward to
an occasion when high precision SF methods applied at second order may be
fruitfully compared with known PN results in the weak field regime.

\section*{References}

\bibliography{BFW_Moriond15.bib}

\end{document}